\documentclass[review]{elsarticle}

\usepackage{lineno,hyperref}
\modulolinenumbers[5]
\usepackage{graphicx}
\usepackage{epstopdf}
\usepackage{color}
\usepackage{soul}
\DeclareGraphicsExtensions{.eps}

\begin{document}

\begin{frontmatter}

\title{Increased stability of CuZrAl metallic glasses prepared by physical vapor deposition}

\author{G.B. Bokas}
\author{L. Zhao}
\author{D. Morgan}
\author{I. Szlufarska\corref{I. Szlufarska}}
\cortext[I. Szlufarska]{Author to whom correspondence should be addressed}
\ead{szlufarska@wisc.edu}
\address{Department of Materials Science and Engineering, University of Wisconsin-Madison, Madison 53706, USA}
\makeatletter
\def\ps@pprintTitle{%
 \let\@oddhead\@empty
 \let\@evenhead\@empty
 \def\@oddfoot{}%
 \let\@evenfoot\@oddfoot}
\makeatother


\begin{abstract}

We carried out molecular dynamics simulations (MD) using realistic empirical potentials for the vapor deposition (VD) of CuZrAl glasses. VD glasses have higher densities and lower potential and inherent structure energies than the melt-quenched glasses for the same alloys. The optimal substrate temperature for the deposition process is 0.625$\times T_\mathrm{g}$. In VD metallic glasses (MGs), the total number of icosahedral like clusters is higher than in the melt-quenched MGs. Surprisingly, the VD glasses have a lower degree of chemical mixing than the melt-quenched glasses. The reason for it is that the melt-quenched MGs can be viewed as frozen liquids, which means that their chemical order is the same as in the liquid state. In contrast, during the formation of the VD MGs, the absence of the liquid state results in the creation of a different chemical order with more Zr-Zr homonuclear bonds compared with the melt-quenched MGs. In order to obtain MGs from melt-quench technique with similarly low energies as in the VD process, the cooling rate during quenching would have to be many orders of magnitude lower than currently accessible to MD simulations. The method proposed in this manuscript is a more efficient way to create MGs by using MD simulations. 

\end{abstract}


\end{frontmatter}

\nolinenumbers

\section{Introduction}

Metallic glasses (MGs) have attracted much attention because of their very good mechanical properties~\cite{Inoue2000, Schuh2007, Shi2006} and their high resistance to corrosion~\cite{ZHUANG2016}. The main drawback of the MGs is their low glass forming ability (GFA). MGs are usually prepared by quenching from liquids with very fast cooling rates in order to avoid crystallization~\cite{Debenedetti2001, Ediger2012, Tang2013}. Recently, it has been found  that increased thermodynamic stability can be achieved in organic glasses prepared by physical vapor deposition (PVD) on a substrate with a controlled temperature~\cite{Mcmahon2007, Dawson2009, Dawson2011, Guo2012}. These PVD grown organic glasses were also found to have higher densities than their quenched counterparts. In addition, their thermodynamic properties (e.g., enthalpy, onset temperature) are equivalent to those of quenched glasses if the latter had been aged for thousands of years. For this reason these PVD-grown organic glasses have been named "ultrastable organic glasses". The best substrate temperature for the creation of ultrastable organic glasses was found experimentally to be around 80-85\% of glass transition temperature ($T_\mathrm{g}$)~\cite{Zhu2010, Ishii2008, Leon-Gutierrez2009, Whitaker2013}. 

Given the success of PVD in organic glasses, it is an interesting question whether this method can be applied to MGs. There have been very few studies focused on creation of ultrastable MGs by PVD. For instance, the authors of Refs.~\cite{Yu2013a, Aji2013} reported successful synthesis of Zr-based MGs and found the optimal substrate temperature to range from 70 to 80\% of the $T_\mathrm{g}$. These VD MGs exhibited higher $T_\mathrm{g}$ than their melt-quenched counterparts and thus they have been referred to as ultrastable MGs. Despite having a higher $T_\mathrm{g}$, the enthalpy of these ultrastable MGs was reported to be higher than the enthalpy of the aged MGs. Thus, it remains an open question whether the ultrastable PVD MG is in a lower energy state than the aged glass.

In order to understand properties of the PVD glasses, molecular dynamics (MD) simulations based on Lennard-Jones (LJ) pair potentials have been carried out and reported in Refs.~\cite{Singh2013, Lyubimov2013, Lin2014, Singh2011, Helfferich2016}. However, pair potentials have a limited applicability to metallic systems where many body effects are important. For instance, pair potentials do not have environmental dependence and therefore they may not be able to capture the energetics of surface atoms correctly.  During the PVD process the surface atoms diffuse faster than bulk atoms and this surface diffusivity has been hypothesized to play an important role in both the final structure and the properties of the PVD glasses~\cite{Yu2013a,Zhu2011, Brian2013a, Zhang2017}. Thus, the goal of this work is to simulate PVD of MGs by MD simulations based on a more realistic empirical potential with multi-body interactions included and compare their structures and properties with those created by  the typical melt-quenched method.

\section{Computational Details}

MD simulations are used to create model films by a process that mimics the PVD. Simulations are performed with the LAMMPS software package~\cite{Plimpton1995} based on the CuZrAl embedded atom method potential~\cite{Cheng2009}. This potential was fitted explicitly to reproduce the potential energy surface calculated by $ab\enskip initio$ calculations of over 600 configurations, including crystalline phases, liquids and MGs. Also, this potential has been validated by comparing the results with both $ab\enskip initio$ and experimental data. In our work, the target composition for the films was selected to be $\mathrm{Cu_{49}Zr_{45}Al_6}$. This particular composition has been found experimentally to form fully amorphous structure and to possess high glass forming ability~\cite{Wang2005}. The substrate, on top of which we grow the MGs, consists of 4,000 LJ atoms and was prepared as a BCC structure. Interactions between LJ atoms of the substrate are relatively strong and the parameters $\epsilon$ and $\sigma$ have been set as 2eV and 2.5$\mathrm{\AA}$, respectively. The cutoff for the LJ atoms interactions is set to 4.5$\mathrm{\AA}$ and the atom mass is set to be $\mathrm{6.64\times 10^{-24}gr}$. The substrate is melted and equilibrated at $T=2,000\mathrm{K}$ and then it is cooled down to various final temperatures. We used the isothermal-isobaric (NPT) ensemble where the temperature and pressure have been controlled with the Nose-Hoover thermostat and barostat, respectively. The pressure is kept at zero. At the end of the quenching procedure, every substrate atom was constrained to its initial position by applying a spring with a force constant $K=1$eV/$\mathrm{\AA}$. The substrate LJ atoms interact with the deposited atoms by weak LJ interactions, i.e., $\epsilon=1$eV and $\sigma=2.25\mathrm{\AA}$.  This weak interaction between the substrate and the deposited atoms  is chosen because the substrate atoms only act as a support for the growing CuZrAl film.
\par Periodic boundary conditions have been applied along the X and Y directions, while along the Z direction (perpendicular to the film) there is a free surface restricted by two repulsive walls, with one below the substrate atoms and the other one at a distance of 150$\mathrm{\AA}$ above the film-substrate interface. The size of the simulation box is 42.05$\mathrm{\AA}$ in both, the X and Y direction. The time step is equal to 1fs for all MD simulations in this study.
\par During the deposition simulation the temperature of the substrate  and  the previously deposited atoms is maintained at the desired substrate temperature $T_\mathrm{s}$ using a canonical-ensemble (NVT). New (gas) atoms are introduced at random positions in the empty space between the film and the repulsive wall with velocities aligned along the Z direction. The magnitude of the initial velocities has been drawn randomly from the Maxwell-Bolzmann distribution at temperatures around $T=2,000\mathrm{K}$. Dynamics of the atoms in the gas phase is modeled in the microcanonical (NVE) ensemble. When a gas atom comes into contact with the solid film, it is first allowed to equilibrate for $\sim 500$ps and after this time it is coupled to the same thermostat as the substrate and the film. The deposition rate in our simulation is $q_\mathrm{d}=0.47\mathrm{m/sec}$. We have performed simulations for substrate temperatures that range from $\mathrm{150K}$ to $\mathrm{1,100K}$ in $\mathrm{50K}$ steps. For every substrate temperature $T_\mathrm{s}$ we have performed 20 independent simulations of PVD. After the deposition of the film, all systems are relaxed for $\mathrm{0.8ns}$ and the data were collected during the next $\mathrm{0.2ns}$ of simulation (the production run). Inherent structure energy of the systems, for a given substrate temperature, is calculated by minimizing the energy of 100 different configurations recorded in the $\mathrm{0.2ns}$ production part of the simulation and taking the average of all these energies from 20 independent systems.
\par We also created melt-quenched systems for comparison of their properties with the PVD-grown films. We performed simulations with cooling rates that varied from $\mathrm{0.1K/ps}$ to $\mathrm{10K/ps}$. Unless otherwise noted, we report the results of simulations with the $\mathrm{0.1K/ps}$ cooling rate. We prepared the melt-quenched systems starting from a cubic cell box consisting of 2,580 atoms, which occupy positions consistent with the B2 structure. The chemical identities of the atoms were randomly assigned to the lattice sites while keeping the average composition  at $\mathrm{Cu_{49}Zr_{45}Al_6}$. Periodic boundary conditions were applied in all three directions. In order to melt the simulated samples, they were heated to and then kept at $\mathrm{2,000K}$ in the constant pressure-constant temperature (NPT) ensemble. Every system was then quenched to $\mathrm{2K}$ in $\mathrm{50K}$ intervals. At every interval we ran NPT simulations at constant $T$ for $\mathrm{15ps}$ to equilibrate the system and subsequently another $\mathrm{15ps}$ to analyze its properties.

\section{Results and Discussion}

\par
\begin{figure}
\centering
\includegraphics[scale=0.40]{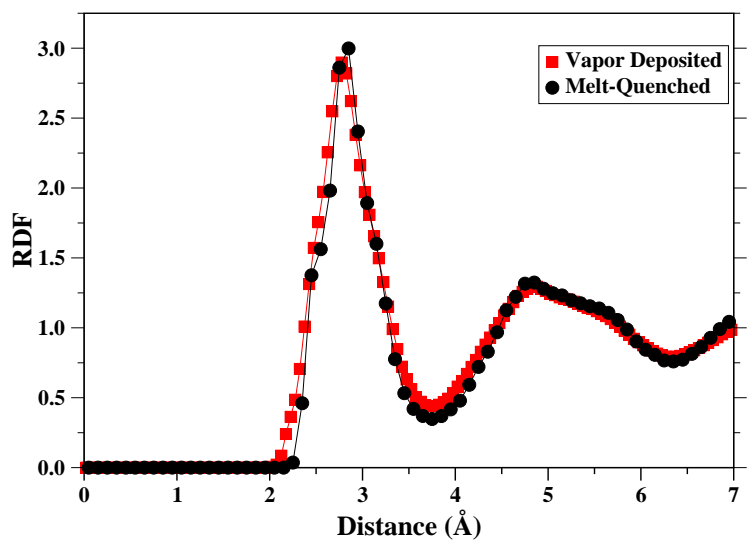}
\caption{(Color online) Comparison of the total radial distribution function of the PVD-grown MGs and melt-quenched MGs prepared at room temperature.}
\label{fig:rdf}
\end{figure}

\begin{figure}
\centering
\includegraphics[scale=0.40]{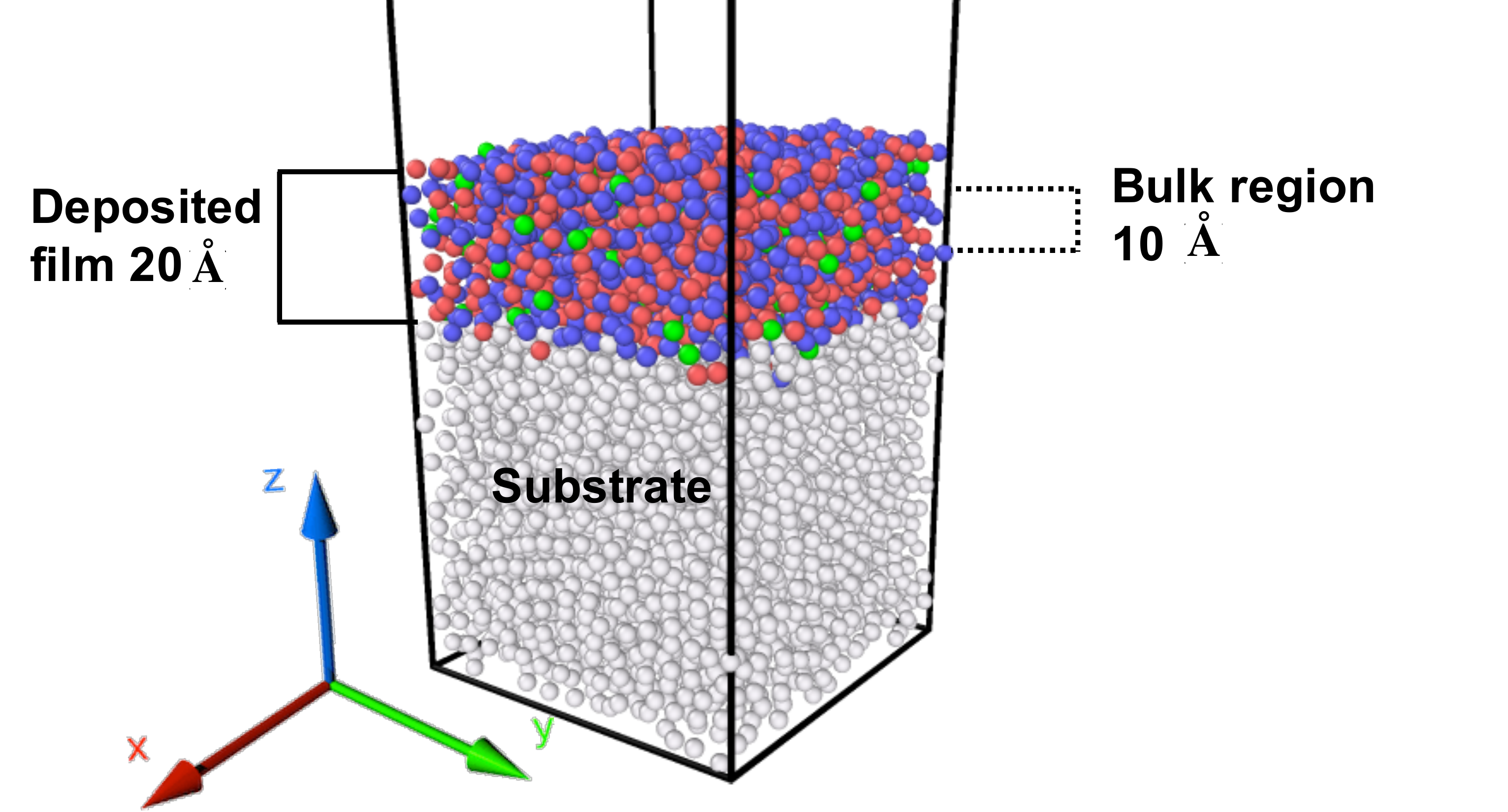}
\caption{(Color online) Geometry of the system for PVD simulations. The thickness of the film at this snapshot is 20$\mathrm{\AA}$. The X and Y dimensions are 42.05$\mathrm{\AA}$. Gray atoms are the substrate atoms; blue, red, and green atoms are the Cu, Zr and Al atoms, respectively. The bulk region corresponds to a plateau in the local potential energy and it extends from 8 to 18$\mathrm{\AA}$.} 
\label{fig:str}
\end{figure}

 In order to confirm that we have indeed formed MGs, we analyzed the radial distribution function (RDF) of the systems for all substrate temperatures. In all samples, we did not find any long-range order (see Fig.~\ref{fig:rdf}) and we observed a small splitting of the second peak of the total RDF (between 4.5 and 6$\mathrm{\AA}$), which is a characteristic feature of MGs and reflects the presence of a medium range order~\cite{Group1977, Zhang2011, Bokas2016}. After the PVD simulations, we calculated the local potential energy and the local composition of the films along the growth direction Z (see Fig.~\ref{fig:str}). The potential energy of the films was found to be lower near the substrate and higher near the free surface of the film, with a plateau in the middle. The thickness of the plateau region is approximately $\mathrm{10\AA}$ and we refer to it as the bulk region. Analysis reported in this manuscript was carried out on the atoms within this central plateau region. Inside this region the composition of the films was also found to be uniform and within $\pm0.5\%$ of the desired composition of $\mathrm{Cu_{49}Zr_{45}Al_{6}}$. As is shown in Fig.~\ref{fig:str}, the total thickness of the film is approximately 20$\mathrm{\AA}$ but the thickness of the bulk region is 10$\mathrm{\AA}$. 
 
\par
\begin{figure}
\centering
\includegraphics[scale=0.50]{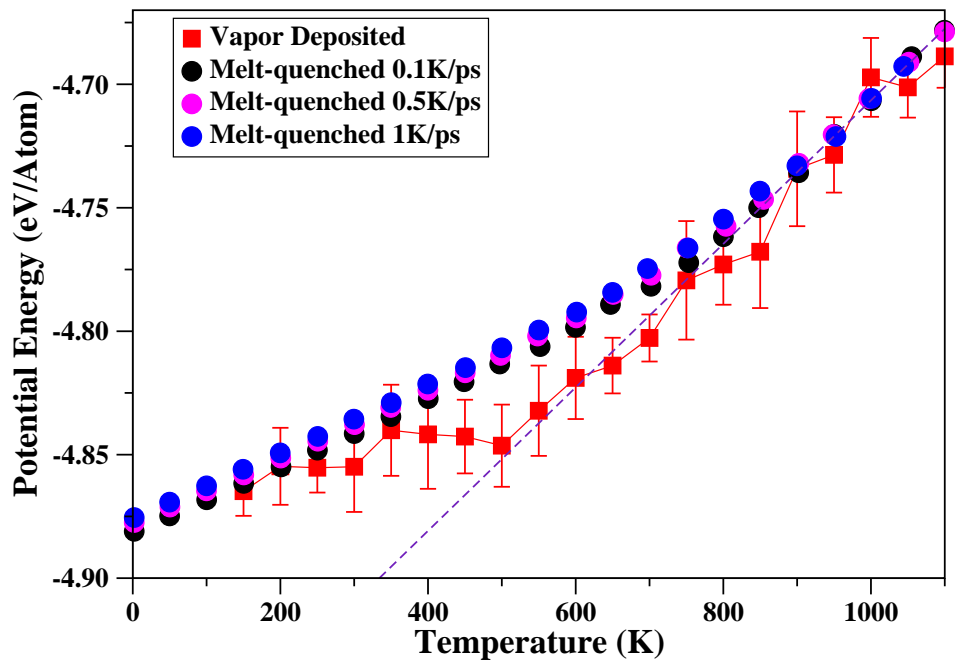}
\caption{(Color online) Potential energy per atom in the PVD glasses (squares) grown at substrate temperature $T_\mathrm{s}$ and of melt-quenched glasses (circles) prepared at different cooling rates. The error bars in the figure correspond to the standard deviation of the potential energy of 20 independent PVD simulations. The melt-quenched glass transition temperature is estimated at $T_\mathrm{g}$=800K.}
\label{fig:potene}
\end{figure}

 In order to find out if the stability of the PVD MGs is increased as compared to the stability of the melt-quenched MGs, in Fig.~\ref{fig:potene} we plot the potential energy per atom at a given temperature. For the PVD MGs the X axis corresponds to $T_\mathrm{s}$ whereas for the melt-quenched glasses the X axis corresponds to the temperature at which we stopped the cooling procedure to equilibrate the system and to calculate its properties. The glass transition temperature, estimated from the change in the slope of the potential energy of the melt-quenched glasses, is $T_\mathrm{g}\approx 800\mathrm{K}$. The extrapolated supercooled liquid line (dashed line in Fig.~\ref{fig:potene}) corresponds to the lowest energy it might be possible for the system to reach without becoming crystalline. As shown in Fig.~\ref{fig:potene}, reducing the cooling rate lowers the energy of the system, but for the cooling rates accessible to MD simulations, the effect is small relative to the energy scale of the plot.

\begin{figure}
\centering
\makebox[\textwidth][c]{\includegraphics[scale=0.50]{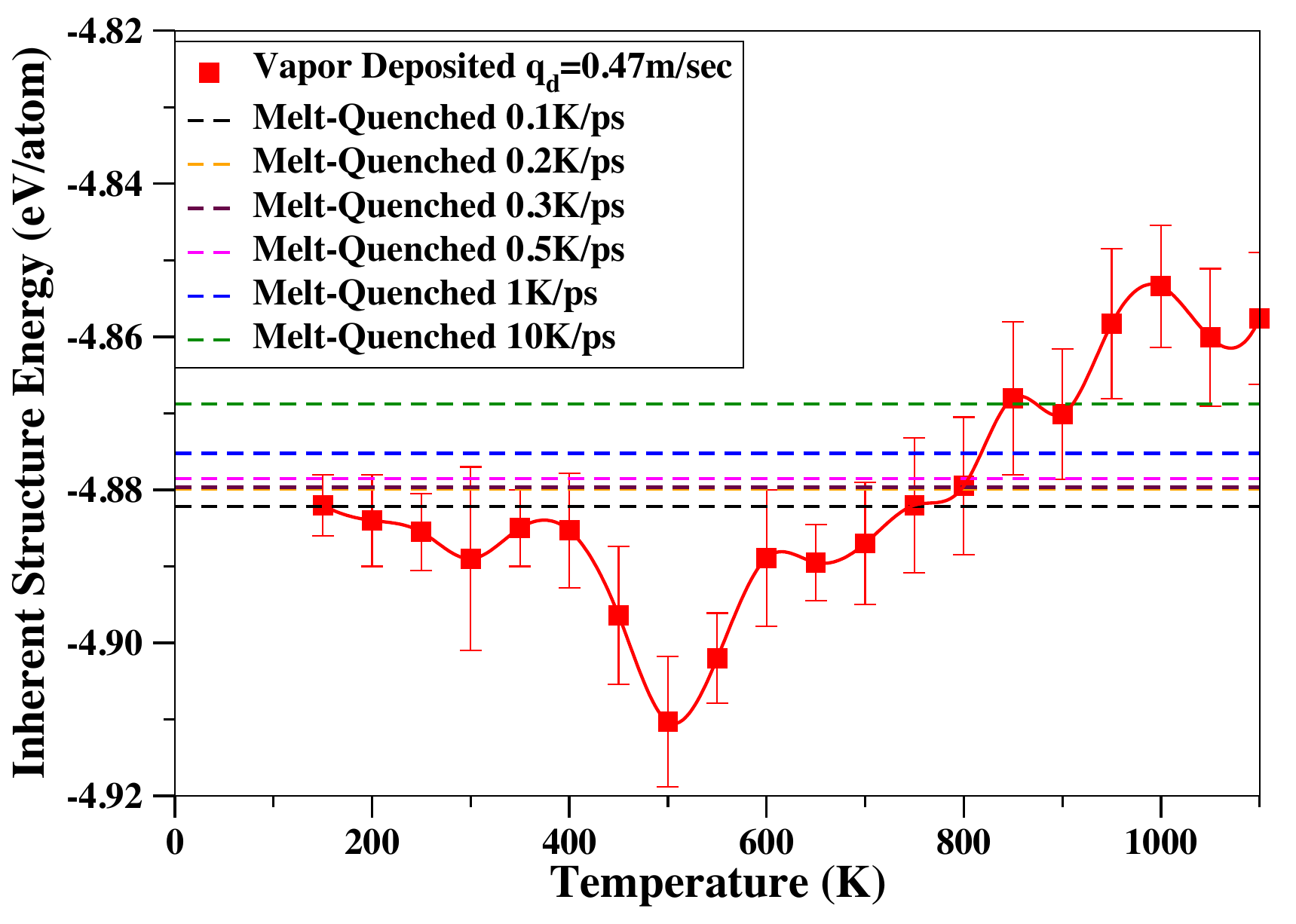}}
\caption{(Color online) Inherent structure energy per atom for PVD films grown at $T_s$ and for melt-quenched glasses prepared at different cooling rates.}
\label{fig:inh}
\end{figure}

The potential energies of the PVD glass and the quenched glass are comparable in two temperature regimes: above $T_\mathrm{g}$ and at low substrate temperatures ($T_\mathrm{s}<400\mathrm{K}$). At low temperatures, the mobility of atoms deposited on the surface is too low and thus the system is not able to find its minimum energy configuration. However, for substrate temperatures between $\mathrm{400K}$ and $\mathrm{700K}$ atoms have enough mobility to find low energy configurations and the potential energy of the PVD glass is lower than that of the melt-quenched glass. Moreover, in this regime the potential energy of the PVD glasses follows the extrapolated supercooled liquid line. This means that the PVD glasses have reached their equilibrium glassy state while the melt-quenched glasses have not.  
In order to find the optimal substrate temperature for the creation of PVD MGs, we calculated the inherent structure energies of the simulated systems (see Fig.~\ref{fig:inh}). The inherent structure energy is determined by minimizing the energy at $T=0\mathrm{K}$ using the steepest-descent method. 

For the PVD MGs the inherent structure energy has a minimum at $T_\mathrm{s}=500\mathrm{K}$, which is 62.5$\%$ of $T_\mathrm{g}$. It has been previously shown by experiments~\cite{Yu2013a} that the optimal substrate temperature in Zr-based MGs is $70-80 \%$ of $T_\mathrm{g}$. In comparison, in organic glasses experimental studies found that the optimal substrate temperature for creating a stable glass is $\mathrm{\sim80-85\%}$ of $T_\mathrm{g}$~\cite{Mcmahon2007, Ramos2011, Kearns2007, Leon-Gutierrez2008}. The reason for this difference between inorganic and organic glasses is not clear. One possible explanation is that the MGs have simpler atomic structures than the atomic structure of the organic-polymeric glasses. Thus, MGs will need shorter time and smaller $T_{\mathrm{sub}}/T_{\mathrm{g}}$ for their atoms to reach their ultrastable state. As the substrate temperature decreases below 400K, the inherent structure energy of the PVD-grown MGs reaches a plateau and is close to that of melt-quenched MGs. On the other hand, for $T_\mathrm{s}>T_\mathrm{g}$,  the PVD-grown MGs have larger inherent structure energies than the melt-quenched MGs, indicating the lower stability of PVD-grown MGs at higher substrate temperatures.

\begin{figure}
\centering
\makebox[\textwidth][c]{\includegraphics[scale=0.60]{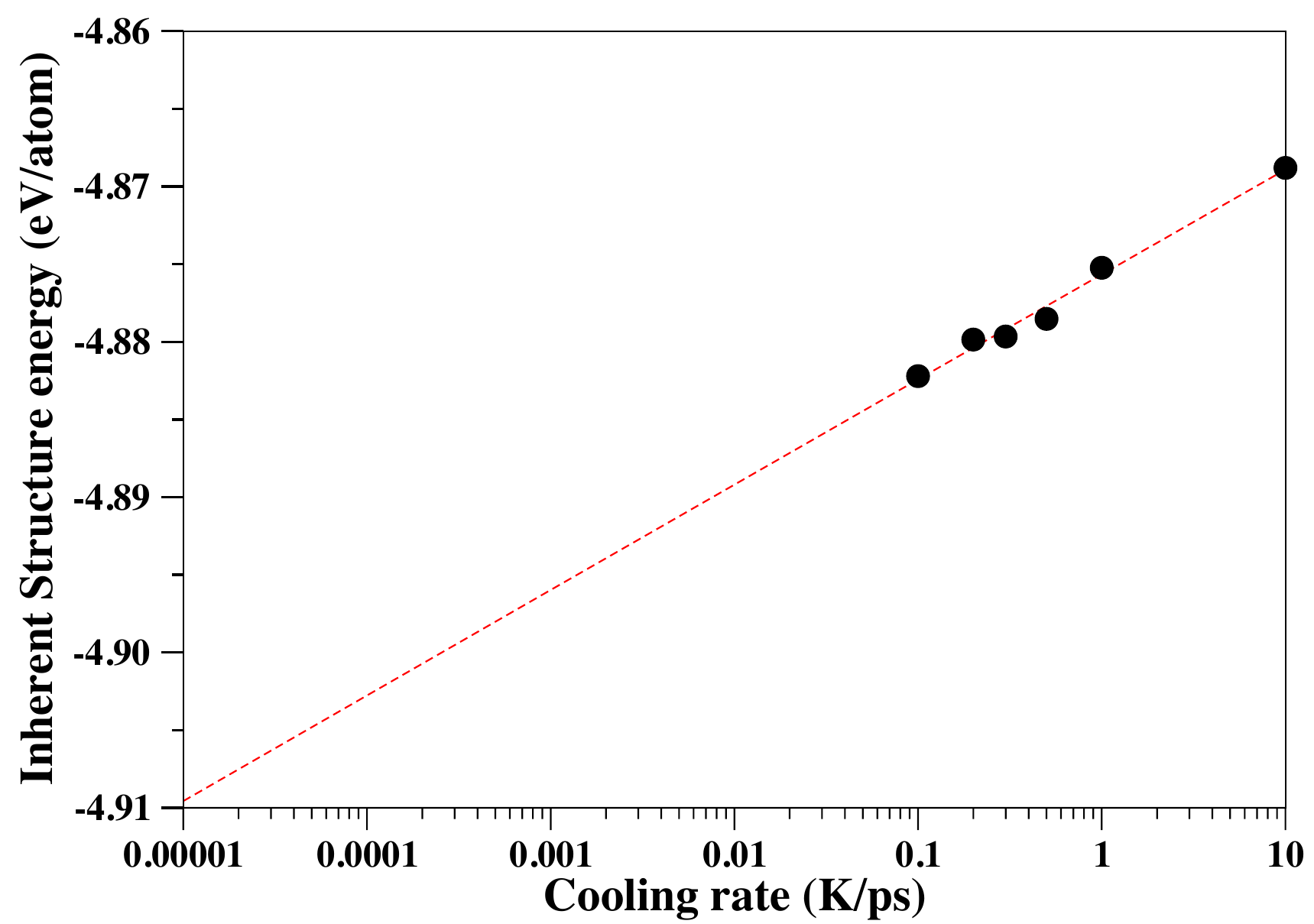}}
\caption{(Color online) Inherent structure energy per atom for melt-quenched glasses prepared at cooling rates $\mathrm{0.1K/ps}$, $\mathrm{0.2K/ps}$, $\mathrm{0.3K/ps}$, $\mathrm{0.5K/ps}$, $\mathrm{1K/ps}$, and $\mathrm{10K/ps}$. Using a logarithmic function to approximate the data, we extrapolated the cooling rate to the inherent structure energy ($\mathrm{-4.91eV}$) of the PVD glass, grown at $T_\mathrm{s}=500\mathrm{K}$. The estimated cooling rate that produces the same inherent structure energy is $\mathrm{10^{-5}K/ps}$.}
\label{fig:fit}
\end{figure}

We found that for melt-quenched glasses, the effect of the cooling rate on the inherent energy is quite significant and, as expected, the lower the cooling rate, the more stable the glass as shown in Fig.~\ref{fig:inh}. One can estimate the cooling rate that would be required to reach the same energy in melt-quench glasses as in the PVD glasses. The lowest inherent structure energy found for PVD glasses is -4.91eV/atom (see Fig.~\ref{fig:inh}). Fig.~\ref{fig:fit} shows corresponding inherent structure energies of the melt-quenched systems versus the cooling rates. Fitting the data with a logarithmic function and extrapolating the fitting function to the inherent structure energy of the PVD glass, we found that the necessary cooling rate has to be at least 4 orders of magnitude slower (i.e., $\mathrm{10^{-5}K/ps}$) than the cooling rate used in this study. Thus, at least in simulations, the PVD method provides an  efficient way to create more stable MGs, which would not be possible using melt-quench simulations. In addition, one should point out that the cooling rate of  $\mathrm{10^{-5}K/ps}$ is achievable in experiments. It has been discussed before~\cite{Lin2014}, that the lower the deposition rate, the lower the energy of the PVD glasses, which means that the PVD glasses created by experiments are going to have even lower energies than our simulated MGs. Therefore, it is likely that, similarly as in simulations, the PVD method in experiments can lead to creation of a more stable MGs than those created by the melt-quenching techniques (i.e., melt spinning).

\par
\begin{figure}
\centering
\includegraphics[scale=0.50]{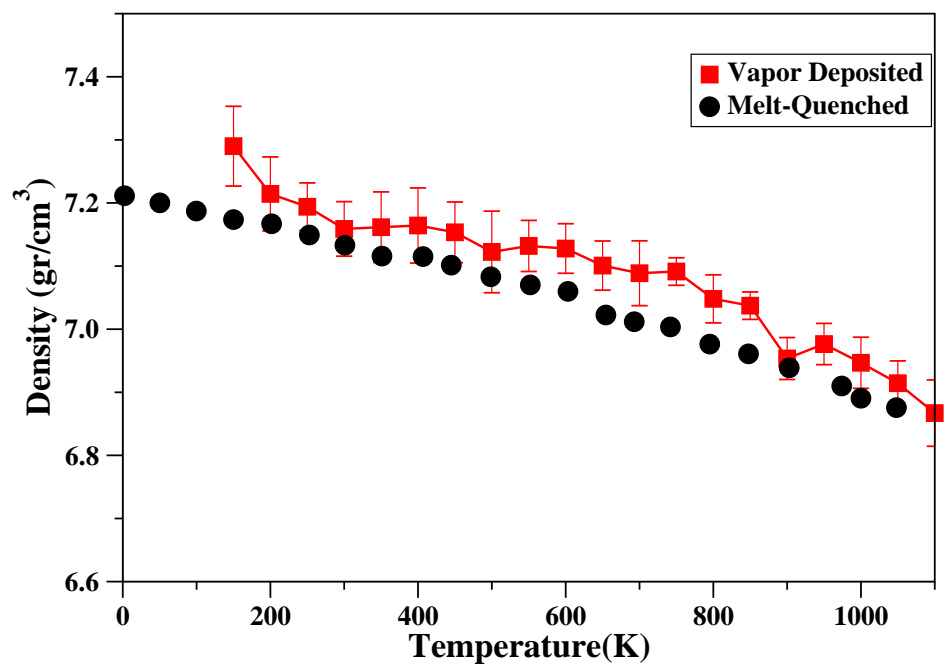}
\caption{(Color online) Density of PVD glasses grown at various substrate temperatures T${_s}$ and melt-quenched glasses during the solidification procedure. The error bars in the figure correspond to the standard deviation of the densities of 20 independent PVD-grown systems.}
\label{fig:den}
\end{figure}

 As described in the introduction, ultrastable organic glasses are characterized by increased thermodynamic stability as well as increased density. To determine if the density is also correlated with the stability  in our PVD-grown MGs, in Fig.~\ref{fig:den} we plot the density of the PVD glasses and the melt-quenched glasses. We find that for $T_\mathrm{s}>\mathrm{850K}$ the two different types of glasses have comparable densities whereas for $T_\mathrm{s}<850K$ the PVD glasses  have a larger density than the melt-quenched glasses by up to $\mathrm{1.5\%}$. The lower potential energy and the higher density of the PVD glasses are in good agreement with experimental findings for the ultrastable organic glasses~\cite{Mcmahon2007, Lin2014}, which shows that the stability in the two classes of materials follow a similar trend.

\par
\begin{figure}
\centering
\includegraphics[scale=0.50]{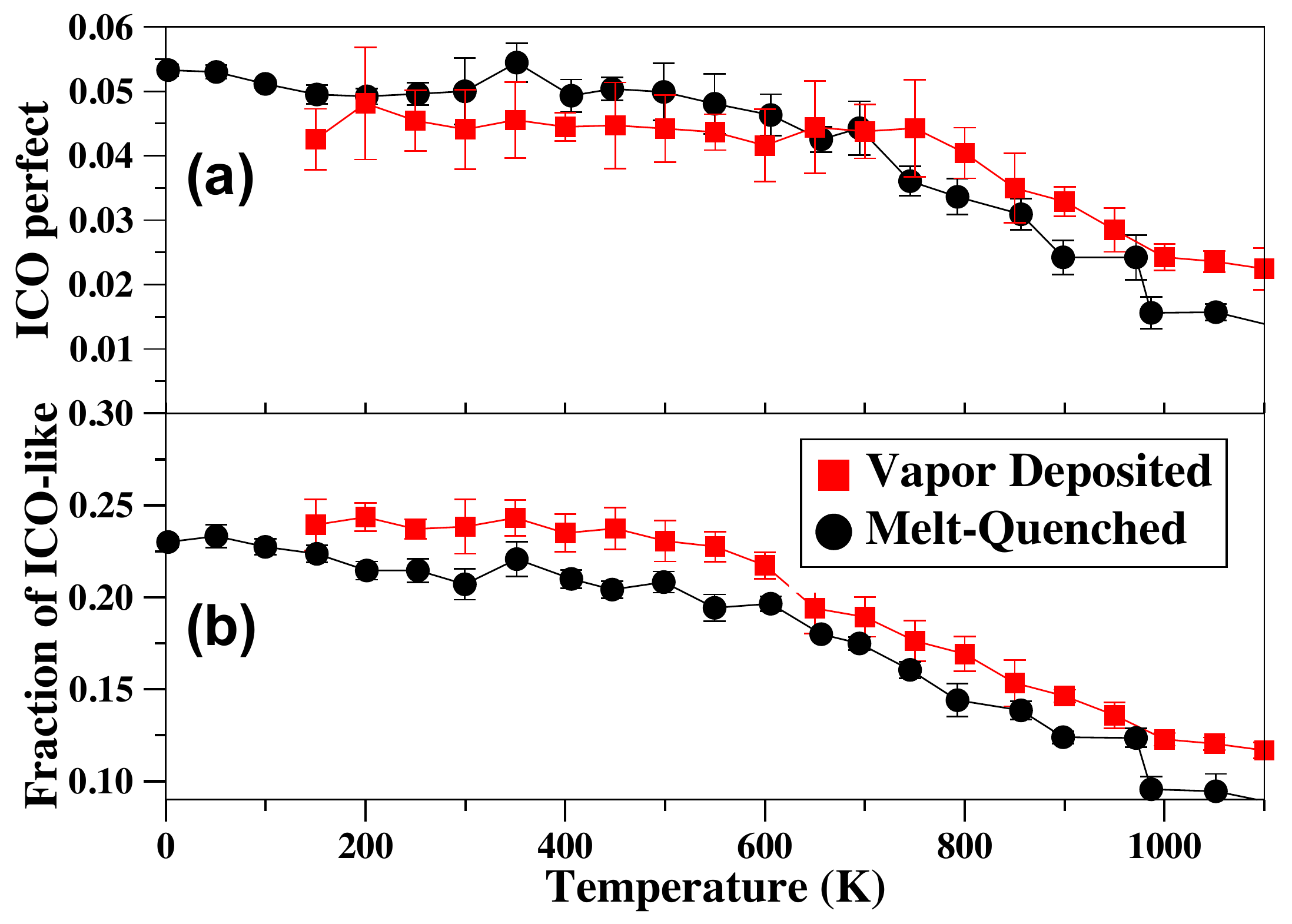}
\caption{(Color online) (a) Comparison of the fractions of the perfect ICO VPs found in the PVD glasses and the melt-quenched glasses. (b) Fractions of ICO-like VPs in the PVD glasses grown at temperature $T_\mathrm{s}$ and melt-quenched glasses.}
\label{fig:like}
\end{figure}

 The differences of the properties between the PVD glasses and melt-quenched glasses have to be reflected in different atomic structures of these two types of glasses. To identify these differences, we first compare the RDF of the PVD glasses and the melt-quenched glasses (Fig.~\ref{fig:rdf}). It can be seen that the two RDFs are almost indistinguishable, which means that the RDF is not sensitive to differences in the structures underlying the observed differences in stability. We have also analyzed the structures using the Voronoi tessellation technique~\cite{Finney1970}. In this technique, each atom of the system is labeled by indices that correspond to the Voronoi polyhedron (VP) centered on that atom. A perfect icosahedral cluster has Voronoi indices equal to (0, 0, 12, 0), which means that it consists only of 12 faces, each with five edges. The icosahedral-like (ICO-like) VPs are those that consist of 10 or more faces with 5 edges each. Glass stability of MGs has been previously proposed to be correlated with the presence of pentagonal faces in VPs~\cite{Cheng2008b, Bokas2016}. By analyzing the structure of the two kinds of glasses at different temperatures we found that the fraction of perfect icosahedral clusters is almost equal in both systems but there is a slight increase in the number of ICO-like clusters in the PVD glasses as compared to the melt-quenched glasses (see Fig.~\ref{fig:like}). The larger number of ICO-like clusters in PVD MGs is consistent with the higher stability (lower energies) of these glasses compared to the melt-quenched glasses since larger number of ICO-like clusters is often associated with a larger stability of glass. However, the number of ICO-like cluster cannot be the only factor controlling stability. If it was, the trend of the fraction of ICO-like VPs would be the same as the trend in the energies of the systems. These trends are not the same in our simulations.

\par Since we are dealing with multi-component alloys, it is also possible the atomic structures of MGs prepared by the two methods are different. To test this hypothesis and to quantify the chemical ordering in the MGs, we calculate the parameter $\alpha_{ij}$, which is analogous to the Warren-Cowley parameter~\cite{warren1990, Han2014, Hong2014, Cheng2011}. Here, we define $\alpha_{ij}=1-\frac{\sum\limits_{j\neq i}Z_{ij}}{\sum\limits_{j}c_{j}Z^i_\mathrm{tot}}$,
where $c_j$ is the concentration of species $j$, $Z^i_\mathrm{tot}$ is the total coordination number of atom $i$ (defined as the total number of atoms in the first shell of the central atom $i$), and $\sum\limits_{j\neq i}Z_{ij}$ is the partial coordination number of atom $i$ (defined as the number of atoms of species $j$ in the first shell of the central atom of species $i$). All these quantities are averaged over the entire system size. A negative value of $\alpha_{ij}$ means that there are more atoms $j$ than $i$ near a central atom $i$ than would be the case in a random solution with the same concentration. A positive value of $\alpha_{ij}$ indicates that there are more atoms $i$ than $j$ near a central atom $i$ than in the case of a random solution. $\alpha_{ij}=0$ corresponds to a random chemical ordering inside the MGs.

\begin{figure}
\centering
\makebox[\textwidth][c]{\includegraphics[scale=0.60]{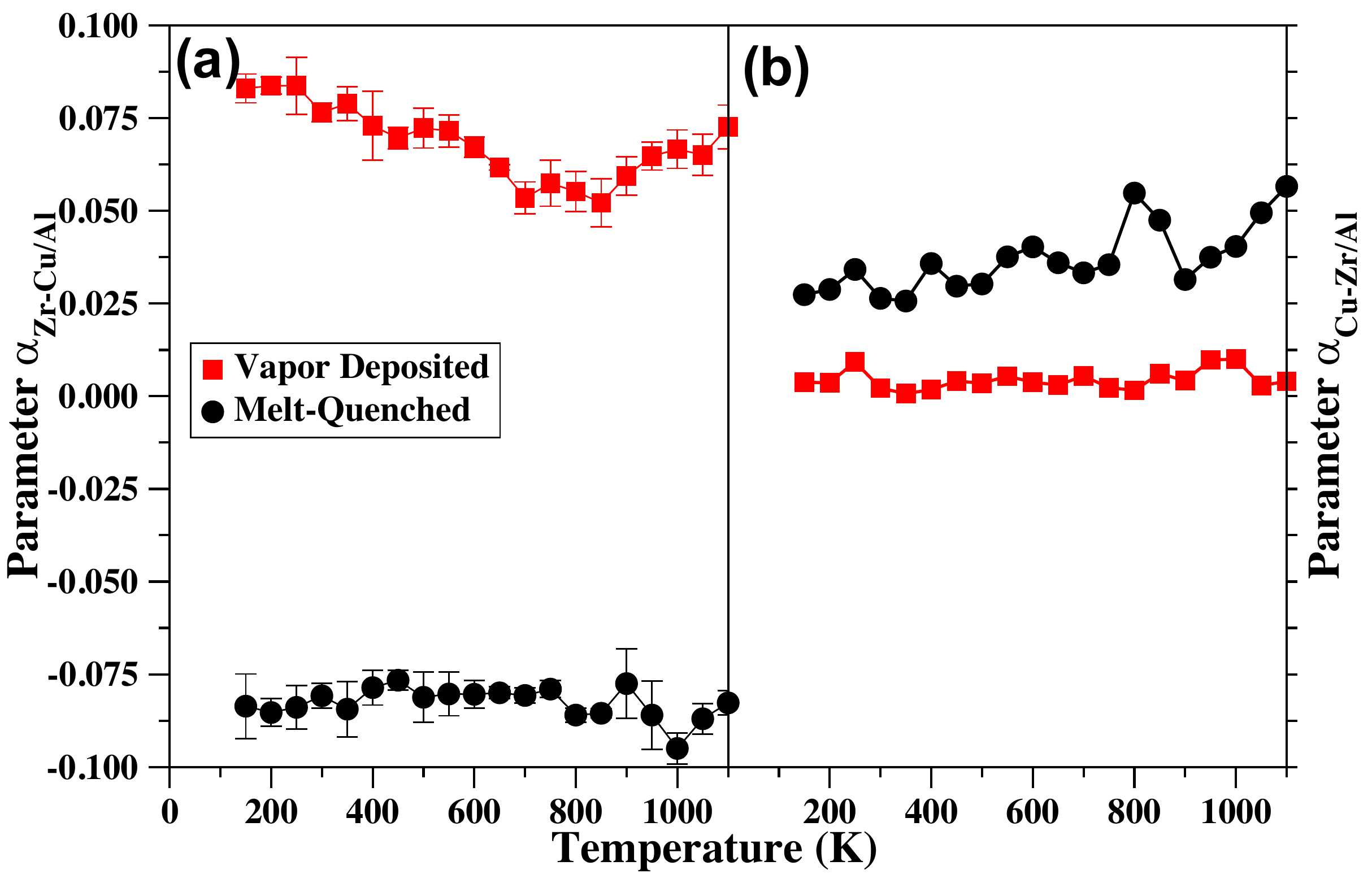}}
\caption{(Color online) (a) The parameter $\alpha_\mathrm{Zr-Cu/Al}$ for Zr-Cu/Al interactions in melt-quenched (circle) and PVD glasses (squared). Positive and negative values of $\alpha_\mathrm{Zr-Cu/Al}$ mean that Zr atoms prefer homonuclear and heteronuclear bonds, respectively. (b) The parameter $\alpha_\mathrm{Cu-Zr/Al}$ for Cu-Zr/Al interactions in melt-quenched (circle) and PVD glasses (squared). $\alpha_\mathrm{Cu-Zr/Al}$ represents the preference of Cu atoms to form homonuclear vs heteronuclear bonds.}
\label{fig:war}
\end{figure}

\par The $\alpha_\mathrm{Zr-Cu/Al}$ parameter is plotted vs temperature in Fig.~\ref{fig:war}a. Since the values of $\alpha_\mathrm{Zr-Cu/Al}$ of the PVD glasses are higher than those of the melt-quenched glasses, it suggests that Zr atoms in the former glasses have a stronger preference to become nearest neighbors compared to the latter ones. The negative values of $\alpha_\mathrm{Zr-Cu/Al}$ measured for the melt-quenched glasses imply that the heteronuclear Zr-Cu and Zr-Al bonds are more favored than Zr-Zr bonds. In other words the number of Cu or Al atoms that surround a Zr atom is larger than would be the case in a randomly packed system. On the other hand, the values of $\alpha_\mathrm{Cu-Zr/Al}$ parameter shown in Fig.~\ref{fig:war}b of the PVD glasses are very close to zero, which means that there is no chemical preference between homonuclear Cu-Cu bonds and heteronuclear Cu-Zr and Cu-Al bonds. In contrast, in the melt-quenched glasses, Cu atoms tend to cluster instead of mixing with Zr/Al atoms, as evidenced by the positive values of the $\alpha_\mathrm{Cu-Zr/Al}$ in Fig.~\ref{fig:war}b. The differences in chemical order between PVD and melt-quenched glasses are likely related to the fact that the atomic structure of the melt-quenched glasses is effectively the structure of a frozen liquid. In contrast, PVD MGs are formed atom by atom without going through the liquid phase. In order to understand whether the increased stability of PVD glasses results from a different chemical ordering (as quantified by the Warren-Cowley parameters $\alpha$), we performed a hybrid Monte Carlo (MC)/ MD simulations to further relax the PVD glasses. Both the MC and MD simulations were performed in the NVT ensemble at the temperatures equal to the temperature of the PVD substrate. In the hybrid simulation scheme, each system was equilibrated using MD for 300ps and the MD run was interrupted every 1 ps to perform an MC attempt to swap randomly selected 15 atoms. Simulations were performed on systems with the substrate temperatures $T_\mathrm{s}=500\mathrm{K}$ and $\mathrm{550K}$, because this is the temperature range where we found the most stable glasses. After the relaxation, the potential energy of the systems slightly decreases (less than 0.006eV/atom), i.e., the glass becomes even more stable. In the meantime, the value of the $\alpha_\mathrm{Zr-Cu/Al}$ parameter decreases as well, moving closer to the value measured for a melt-quench glass. This result means that the PVD glass has a lower energy than the melt-quench glass, despite the fact that the PVD glass has a chemical order that is energetically less favorable. Chemical order is therefore not responsible for the larger stability of the PVD glass.


\section{Conclusions}

Realistic empirical potentials that include many-body interactions for metals have been used to study properties of PVD MGs and to compare them with those of melt-quenched MGs. We found that there is an optimal substrate temperature (62.5$\%$ of $T_\mathrm{g}$) for PVD that produces MGs with the lowest inherent energy and the highest density. In addition, PVD glass is found to have a higher degree of chemical ordering. Increased stability of PVD glass is correlated to some extent with a higher number of icosahedral-like Voronoi polyhedra and with a higher density in this material, but none of the properties investigated in this study (i.e., density, icosahedral-like fraction, chemical ordering) was found to have a perfect correlation with the increased stability found in the PVD glasses. It is still an open question if there is any detectable structural signature correlated with this increased stability. One possible explanation might be that multiple features (such as density, ICO fraction, chemical order, etc.) in PVD contribute collectively to its increased stability.
\par In the melt-quenched Cu-Zr metallic glasses, chemical bonding between unlike atoms are favored, which may be related to the fact that these MGs have a memory of the atomic structure of the liquid state. In contrast, PVD MGs are formed atom by atom and the absence of the liquid state may result at different chemical ordering compared to the melt-quenched glasses. Finally, to create melt-quenched glasses with the same energies as PVD glasses the cooling rate should be at least 4 order of magnitude slower than the cooling rate used in this study.

\section*{Acknowledgments}

This research was primarily supported by NSF through the University of Wisconsin Materials Research Science and Engineering Center (DMR-1720415). The authors gratefully acknowledge helpful discussions with Prof. J. de Pablo from University of Chicago.

\section*{References}

\bibliography{library}

\end{document}